
\magnification 1200
\magnification 1200
\baselineskip=16pt
\overfullrule 0pt
\nopagenumbers
\headline={\hfill \tenrm\folio\ }
\parskip 10pt plus 1pt minus 1pt
\hsize 16 true cm \vsize 24 true cm
\def\relatif{\ \hbox{{\rm Z}\kern-.4em\hbox{\rm Z}}}
\def\reel{\ \hbox{{\rm R}\kern-1em\hbox{{\rm I}}}}
\def\souligne#1{$\underline{\smash{\hbox{#1}}}$}
\def\1op{\hbox{{\rm 1}\kern -0.23em\hbox{{\rm I}}}}
\def\laeq{\ \raise .3ex\hbox{{\rm {$<$}}\kern-.8em\lower 1.3ex\hbox{{\rm
{$\sim$}}}}\ }
\def\gaeq{\ \raise .3ex\hbox{{\rm {$>$}}\kern-.8em\lower 1.3ex\hbox{{\rm
{$\sim$}}}}\ }

\ \
\vskip 2 cm
\centerline{\bf
 Electroweak
 Bubbles and Sphalerons}
\vskip 1.5 true cm

\centerline {\bf Yves Brihaye}
\centerline {Facult\'e des Sciences, Universit\'e de Mons-Hainaut
             B-7000 Mons Belgium}
\vskip 0.5 cm
\centerline {\bf Jutta Kunz}
\centerline {Instituut voor Theoretische Fysica, Rijksuniversiteit
Utrecht, The Netherlands}
\centerline {and}
\centerline {FB Physik, Postfach 2503, Universit\"at Oldenburg, Germany}
\vskip 3 cm
\centerline{\bf Abstract}

We consider non-perturbative solutions of
the Weinberg-Salam model at finite temperature.
We employ an effective temperature-dependent potential
yielding a first order phase transition.
In the region of the phase transition,
there exist two kinds of static, spherically symmetric solutions:
sphalerons and bubbles.
We analyze these solutions as functions of temperature.
We consider the most general spherically symmetric fluctuations
about the two solutions and construct the discrete modes
in the region of the phase transition.
Sphalerons and bubbles both possess a single unstable mode.
We present simple approximation formulae for these levels.

\vfill
\noindent {Mons-Preprint} \hfill\break
\noindent {Utrecht-Preprint THU-93/09} \hfill\break
\noindent {April 1993}
\vfill \eject

\noindent {\bf 1) \souligne {Introduction}}

The Weinberg-Salam model describes the electroweak interactions
in the perturbative regime with remarkable accuracy.
The model also predicts non-perturbative processes,
such as instanton or sphaleron induced transitions,
leading to baryon number violation.
The suggestion [1] that the baryon asymmetry of the universe
may possibly be explained within the standard model
or its variants, recently created much interest
in electroweak sphalerons and bubbles,
non-perturbative solutions of the model.

Sphalerons [2] are relevant for baryon number violating processes
in the early universe [3-5].
At a given temperature
the rate of baryon number violation is largely
determined by the three-dimensional action of the sphaleron
solution, which enters in the Boltzmann factor,
while the preexponential factor contains the frequencies of
small oscillations about the sphaleron [3].
If the rate is large enough, the baryon asymmetry
of the universe may be generated during
the electroweak phase transition.

However, for the creation of the baryon asymmetry of the universe
during a first order electroweak phase transition,
the details of this phase transition turn out to be important and
need to be better understood [6-11].
The first order phase transition proceeds
via the creation of bubbles of the new broken phase
embedded in the surrounding unbroken phase.
The rate for the creation of bubbles is largely
determined by the three-dimensional action of the bubble
solution, which enters in the Boltzmann factor,
while the preexponential factor contains the frequencies of
small oscillations about the bubble [7].

Bubbles and sphalerons both arise as non-perturbative
solutions of the finite temperature electroweak field theory.
In the effective potential approach [12],
the ``Mexican hat'' Higgs potential of the electroweak model
is supplemented by temperature dependent terms.
It is the presence of a term cubic in the Higgs field
which renders the phase transition first order.

The sphaleron of the electroweak theory, present at zero temperature,
persists at finite temperature up to the region
of the phase transition [13].
In contrast, the bubble solution exists only at finite temperature,
in the region of the phase transition [7].
Here we study these non-perturbative solutions of the temperature
dependent model in the region of the phase transition.
In the first part of the paper we discuss the properties of these
solutions, e.~g.~we address their three-dimensional action,
their shape and their size.
In the second part of the paper we analyze the discrete modes
about these solutions.
We consider the most general spherically symmetric fluctuations,
diagonalize the quadratic form in these fluctuations and
compute the eigenvalues of the normal modes.
We find that both sphalerons and
bubbles possess a single unstable mode.
For the negative eigenvalues we present approximation
formulae.
We relate the discrete modes of the bubble to those of
the antikink.

\vskip 0.5 true cm

\noindent {\bf 2) \souligne {Lagrangian and effective potential}}

We consider the electroweak theory in the
limit of zero mixing angle.
The bosonic part of the Lagrangian reads
$${\cal L} = -{1\over 4} F^a_{\mu\nu} F^{a\mu\nu}
  + (D_{\mu}\Phi)^\dagger (D^{\mu}\Phi)-V(\phi,T) \ , \eqno(1)$$
$$\phi(x) = \sqrt {2} (\Phi^{\dagger} \Phi)^{1\over 2} \ , $$
with the covariant derivative and field strength tensor
$$D_{\mu}\Phi = (\partial_{\mu}-i{g\over 2}
  \tau^aA^a_{\mu})\Phi\ , \eqno(2)$$
$$F^a_{\mu\nu} = \partial_{\mu} A^a_{\nu} - \partial_{\nu}A^a_{\mu}
  + g \epsilon_{abc} A^b_{\mu} A^c_{\nu}\ . \eqno(3)$$
$V(\phi,T)$ denotes the temperature dependent effective potential
[4,9-11]
$$V(\phi,T) = {\lambda\over 4} \phi^4
   - {\lambda \over 2} v^2 \phi^2
  +{\gamma\over 2} T^2\phi^2-\delta T\phi^3 + c \ ,  \eqno(4)$$
$$\gamma = {2M^2_W+M^2_Z+2M^2_t\over {4v^2}} \ , \eqno(5)$$
$$\delta = {2M^3_W+M^3_Z\over {4\pi v^3}} \ , \eqno(6)$$
$v$ is the Higgs field expectation value at zero temperature,
and $M_W$, $M_Z$, $M_t$ denote the
masses of the $W^{\pm}$ bosons, the $Z^0$ boson and
the top quark, and $c$ is an adjustable constant.

The effective potential (4)
gives rise to a first order phase transition, characterized by three
critical temperatures. For low values of
the temperature $T$, the minimum of $V(\phi,T)$
is attained at some value
$\langle \phi(T) \rangle \not= 0$, while $\phi=0$
corresponds to a local maximum of $V(\phi,T)$.
At $T=T_c$ with
$$ T^2_c = {\lambda v^2\over {\gamma}}\eqno(7)$$
the second derivative of the potential at the origin changes sign.
For $T>T_c$ the extremum $\phi=0$ turns
into a local minimum separated from the absolute minimum
$\langle \phi(T) \rangle$
by a small energy barrier.
The two minima become degenerate at $T=T_b$ with
$$ T^2_b = {{T^2_c}\over{1-{{2\delta^2}\over{\lambda\gamma}}}}
\ . \eqno(8)$$
For $T>T_b$ the extremum $\phi=0$ constitutes the absolute minimum
of the potential.
Finally at $T=T_a$ the local minimum disappears.

To study temperature dependent quantities
in the region of the phase transition
it is convenient
to represent the temperature via the variable $\xi$ [10]
$$\xi = {\lambda \gamma\over {\delta^2}} \Bigl(1-({T_c\over T})^2\Bigr)
\ . \eqno(9)$$
$T_c$, $T_b$, and $T_a$ then correspond
to $\xi = 0$, $\xi=2$, and $\xi={9\over 4}$, respectively.

Throughout the paper we employ for the masses the values
$M_W = 80$ GeV, $M_Z=92$ GeV, $M_t = 120$ GeV and
$M_H = 45$ GeV,
and the vacuum expectation value for the Higgs field
$\langle \phi(0)\rangle = v = 246$ GeV,
leading to the critical temperatures
$$  T_c = 0.284 v\ ,
\ \ T_b = 0.292 v\ ,
\ \ T_a = 0.293 v\ . \eqno(10)$$

\vskip 0.5 true cm
\noindent {\bf 3) \souligne {Non-perturbative solutions}}

In order to construct non-perturbative solutions
of the effective lagrangian (1)
we choose the gauge $A^a_0 = 0$ and we employ for the fields the static,
spherically symmetric ansatz
$$\Phi(\vec r) = {v\over {\sqrt {2}}} L(r) \pmatrix {0\cr 1\cr}\ ,
\ \ \ A^a_i(\vec r) = {G(r)\over {gr^2}} \epsilon_{iba}r_b
\ , \eqno(11)$$
where $L(r)$ and $G(r)$ are radial functions.
The three-dimensional effective action is then given by the functional
$$  S(T) = {4 \pi M_W \over {g^2}} \int dx x^2 \sigma(x)
\ , \eqno(12a)$$
with the action density
$$\sigma(x) =
{G^2(G-2)^2\over {2x^2}}
+ ({dG\over {dx}})^2 + 2x^2({dL\over {dx}})^2+G^2L^2
+ {16\lambda\over {g^2}} x^2 V(L,T)
\ , \eqno(12b)$$
the effective potential
$$ V(L,T) = {1\over 4}
 \Bigl(L^4-2L^2+{2\gamma T^2 \over {\lambda v^2}}L^2
 - {4\delta T\over {\lambda v}} L^3+c \Bigr)
\ , \eqno (12c)$$
and the dimensionless variable
$$x = M_Wr\ ,\ \ \ M_W = {1\over 2}gv \ . $$
The constant $c$ in the effective potential has to be adjusted
such that the integrand (12b) approaches zero asymptotically for the
non-perturbative solution considered.

\noindent {\souligne {Sphalerons}}

The sphaleron, a well-known non-perturbative solution
of the effective action (12) for $T=0$, can be
continued to finite temperatures, $0\leq T<T_a$.
The sphaleron functions satisfy the boundary conditions
$$G(0) = 2\ , \ \ \ \ \ G(\infty) = 0\ ,\eqno(13a)$$
$$\ \ \ \ \ L(0) = 0\ ,
  \ \ \ \ \ L(\infty) = {\langle \phi(T) \rangle \over v}
\ , \eqno(13b)$$
where $\langle \phi(T)\rangle$ denotes
the non-trivial minimum of $V(\phi,T)$.
The profiles of $G(x)$ and $L(x)$ are presented in Fig.~1
for $T=T_c$ and $T=0$.
The action density (12b) of the sphaleron
is presented for the temperatures
$T=T_b$, $T=T_c$ and $T=0$ in Fig.~2.
Both figures indicate that the sphaleron radius
increases from $r\approx 3 M_W^{-1}$
at $T=0$ to $r\approx 9 M_W^{-1}$ for $T=T_b$.

In Fig.~3 we exhibit the three-dimensional action of the sphaleron
as a function of $\xi$ in the range of the phase transition.
The three-dimensional action of the sphaleron
follows an approximate scaling law,
$$S_{\rm sp}(T) = S_{\rm sp}(0)
  {\langle \phi(T) \rangle \over {\langle \phi(0)\rangle}}
\ . \eqno(14)$$
For an effective potential without the cubic term
this scaling law is exact.
$S_{\rm sp}(T)$ is well approximated by the scaling formula (14) [13].
The values predicted by the scaling formula
exceed the exact numerical values only by a few percent,
from 3\% at $T=T_c$ up to 6\% at $T=T_b$.
For comparison the values obtained from the scaling formula (14)
are also shown in Fig.~3.
Since the scaling formula works so well,
we have introduced a scaling factor
$$\Lambda= {\langle \phi(T) \rangle \over {\langle \phi(T_b)\rangle}}$$
in Fig.~2 for the coordinate $x$ and for the action density
$\sigma(x)$. In the case of exact scaling, the three curves
would be equal.

\noindent {\souligne {Bubbles}}

In the range of the phase transition
$T_c<T<T_b$, the equations of motion
obtained from the functional (12) admit another
non-perturbative solution, the bubble.
It is through the formation of bubbles,
that the first order electroweak phase transition proceeds.

For the bubble solution, the gauge field consistently vanishes,
$G(x) = 0$, and the equation for the Higgs field function $L(x)$ reads
$${d^2 L \over {d x^2}}
 + {2\over x} {d L\over {d x}} =
  {4\lambda\over {g^2}} \Bigl[
   L^3 - L\Bigl(1-{\gamma T^2\over {\lambda v^2}}\Bigr)
 - L^2 \Bigl({3\delta T \over {\lambda v}}\Bigr)\Bigr]
\ . \eqno(15)$$
Choosing $c=0$,
a regular solution with finite three-dimensional action is obtained,
where the function $L(x)$ satisfies the boundary conditions
$${\partial L\over {\partial x}}(0) = 0\ ,\ \ \ \ \ L(\infty) = 0
\ , \eqno(16)$$
to be constrasted with boundary condition (13b).
Note, that Derrick's theorem,
forbidding the existence
of static solutions of finite energy of
a scalar theory in three spatial dimensions,
applies only to a positive definite potential.

The phase transition occurs in the
temperature interval $1.3 < \xi < 1.8$ [10,11].
The bubble function $L(x)$
is presented in Fig.~4 for the temperatures
$\xi=1.3$, $\xi=1.8$ and $\xi=1.9$.
Only when $\xi > 1.6$ the bubble starts to develop an interior
region, where $L(x)$ is approximately constant.
When $\xi=1.8$ the size of this interior region is only
about the size of the wall,
and roughly $60M_W^{-1}$ wide.
When $\xi=1.9$
the interior region is only about twice the size of the wall.
Increasing the temperature beyond $\xi=1.9$ leads
to a rapid increase in the size of the interior region,
while the wall approaches its asymptotic shape,
an antikink solution [10,11]
$$L(x)= {{\delta T_b}\over{\lambda v}}
  \Bigl[ 1 - {\rm tanh} \Bigl( {1\over{\sqrt{2 \lambda}}}
 {{\delta T_b}\over{M_W}} (x-X) \Bigr) \Bigr]
\ , \eqno(17)$$
where $X$ denotes the location of the antikink,
which in the limit $\xi=2$ moves towards infinity.

Throughout the region $1.3 < \xi < 1.8$
the thin wall approximation cannot be employed.
This is also clear from Fig.~5,
where we show the action densities
for the temperatures $\xi=1.3$, $\xi=1.8$ and $\xi=1.9$.
For the potential considered,
the thin wall approximation is valid only for $\xi$ very close to 2.

In Fig.~6 we show the quantity $X$, defined as the position
where the derivative of the Higgs field function $L'(x)$
assumes its maximal absolute value
$$X : \ \ L'(X)<0 \ , \ \ L''(X)=0 \ . \eqno(18)$$
This value of $X$ we employ as a definition for the location
of the bubble wall. It is consistent with the use of $X$
in eq.~(17) in the limit $\xi=2$.
Approaching this limit we observe a rapid increase in $X$.
Here $X$ can be interpreted as the bubble radius.

Fig.~(6) also exhibits the quantity $D$
$$D= -{L(0)\over{L'(X)}} \ , \eqno(19)$$
which we interpret as a measure for the size of the wall.
In the limit $\xi=2$ $D$ smoothly approaches
the corresponding analytical value of the antikink
$$D= {{2 M_H M_W}\over{\delta T_b v}}\ . $$

The value of the Higgs field function
at the origin, $L(0)$,
is shown in Fig.~7 as a function of $\xi$.
$L(0)$ starts from zero at $\xi=0$,
and increases
with increasing $\xi$ until about $\xi=1.7$,
where it gets close to $\langle \phi(T) \rangle$.
Then it decreases along with $\langle \phi(T) \rangle$,
approaching $\langle \phi(T) \rangle$ from below.

The three-dimensional action of the bubble
obtained from eqs.~(12) with $c=0$
is presented as a function of $\xi$ in Fig.~3.
We find good agreement with the semi-analytic
formula presented by Adams [14].
In the limit $\xi = 2$,
the action diverges, and the bubble ceases to exist.
\vskip 0.5 true cm

\noindent {\bf 4) \souligne {Normal modes}}

Sphalerons and bubbles are extrema of the
effective action (12).
None of them is a stable minimum.
To exhibit their instabilities
one needs to consider general fluctuations about
the extremal configurations
and to construct the parts of the action functional quadratic in these
fluctuations.
The quadratic parts then represent operator equations,
whose discrete negative eigenvalues correspond to the
unstable modes of the non-perturbative solutions.

By construction, the non-perturbative solutions discussed above
are invariant under $SO(3)_d$, a diagonal subgroup of
$SU(2)_I \times S0(3)_R$ (i.~e.~the group of weak isospin and
the group of spatial rotations).
General fluctuations about the
non-perturbative solutions can be decomposed
according to the irreducible representations of $SO(3)_d$,
whose contributions decouple in the quadratic form.
Here, we consider only the sector of spherically
symmetric fluctuations, where the instabilities of
the zero-temperature sphaleron are known to reside
as well as the unstable breathing mode of the bubble.

\noindent {\souligne {Normal modes about the sphalerons}}

The most general spherically symmetric fluctuations about
the sphaleron consists of five radial functions
$$\Phi(\vec x) = {v\over {\sqrt {2}}}
      \Bigl( (L(x) + {\psi(x)\over {\sqrt{2}x}}) \1op \
    + i {\chi(x)\over{\sqrt{2}x}}
     ({\vec x \cdot \vec \tau\over x}) \Bigr)  \pmatrix {0\cr 1\cr}
\ , \eqno(20)$$
$$A^a_i (\vec x) = {M_W\over {gx}}
      \Bigl[ \Bigl(G(x) + \psi_G(x) \Bigr)
      \epsilon_{iba} \hat x_b
    + \psi_H(x) (\delta_{ia}-\hat x_i \hat x_a)
    + \sqrt {2}\psi_K (x) \hat x_i\hat x_a \Bigr] \ . \eqno(21)$$
The fluctuations $\psi$ and $\psi_G$ decouple
in the quad\-ra\-tic form from the fluctuations
$\chi$, $\psi_H$ and $\psi_K$ [15],
yielding a system of two equations (2-channel)
and a system of 3 equations (3-channel), respectively,
$$\eqalign {&\psi''_G = \Bigl (L^2+{3(G-1)^2-1\over {x^2}}
  - \omega^2\Bigr ) \psi_G + {\sqrt {2} LG\over x} \psi\ , \cr
  &\psi'' =  {\sqrt {2} LG\over x} \psi_G + \Bigl (
    {G^2\over {2x^2}}
   + {{4\lambda}\over{g^2}} {d^2 V \over {d L^2}}
   - \omega^2 \Bigr) \psi\ , \cr}
 \eqno(22)$$
$$\eqalign {&\psi'_K = -{1\over x} \psi_K - {\sqrt {2}(G-1)\over x}
  \psi_H + L\chi \ , \cr
  &\psi''_H = {2\sqrt {2}(G-1-xG')\over {x^2}} \psi_K
  + \Bigl (L^2+{3(G-1)^2-1\over {x^2}} - \omega^2\Bigr ) \psi_H
  - {\sqrt {2}LG\over x}\chi \ , \cr
&\chi'' = 2L'\psi_K
  - {\sqrt {2}LG\over x} \psi_H
  + \Bigl ({(G-2)^2\over {2x^2}} + L^2
  + {{4\lambda}\over{g^2}} {1\over L} {d V \over {d L}}
  - \omega^2\Bigr ) \chi
\ . \cr} \eqno(23)$$

For $T=0$, the negative mode of the sphaleron, present in
the 3-channel [15], is unique
for $M_H < 12 M_W$ [16,17], while new directions of
instability develop for $M_H \ge 12 M_W$. The 2-channel
possesses a single discrete positive mode for
$0\leq M_H\leq 1.5 M_W$ [17].

Also for $T>0$
the sphaleron possesses (at $M_H=45$ GeV)
a unique negative mode in the 3-channel, shown in Fig.~8.
Analogous to the approximate scaling behaviour (14)
of the three-dimensional action
we find an approximate scaling behaviour of the eigenvalue
$\omega^2$ of the negative mode
$$\omega^2_{\rm sp}(T) = \omega^2_{\rm sp}(0)
  {\langle \phi(T)\rangle^2\over {\langle \phi(0)\rangle^2 }}
\ , \eqno(24)$$
which again is exact for an effective potential without cubic term.
Using $\omega^2(0) = -1.9$, the deviation of
the numerical values of $\omega^2(T)$
from the approximate scaling values (24) is
on the order of 10\% at $T=T_c$ (i.~e.~$\xi=0$)
and increases to about 20\% at $T=T_b$ (i.~e.~$\xi = 2$).

\noindent {\souligne {Normal modes about the bubbles}}

For the discrete modes of the bubble
the 2-channel eqs.~(22) is the relevant channel.
For the bubble the two equations
of the 2-channel decouple, since $G(x)=0$.
The equation for the fluctuation $\psi_G$
has a repulsive potential
and therefore no normalizable discrete modes.

The equation for the fluctuation $\psi$ reduces to
$$ -\psi'' + {4\lambda\over {g^2}}
   \Bigl( 3L^2-{6\delta T \over {\lambda v}} L
         - 1 + {\gamma T^2\over {\lambda v^2}}\Bigr) \psi =
   \omega^2\psi \ . \eqno(25)$$
The potential in eq.~(25) is ``deep enough''
as to allow for a state with negative
eigenvalue (i.~e.~$\omega^2 < 0$).
This eigenmode represents
the unstable breathing mode of the bubble,
which lets bubbles smaller than the critical bubble
shrink and larger bubbles grow.
The negative eigenvalue $\omega^2$ is presented in Fig.~9.
It starts from zero at $\xi=0$,
where the bubble solution appears, and ends at zero at $\xi=2$,
where the bubble solution approaches the antikink eq.~(17),
which also has a zero eigenvalue.
For $\xi \rightarrow 2$ the mode of the bubble indeed
approaches the zero mode of the antikink
$$\psi(x)= {1\over{{\rm cosh}^2\Bigl( {1\over{\sqrt{2 \lambda}}}
 {{\delta T_b}\over{M_W}} (x-X) \Bigr) }} \ , \ \ \ \
  \omega^2=0 \ . $$

Fig.~9 also shows approximate values
for the eigenvalue $\omega^2$,
obtained from the formula
$$ \omega^2 = -{2\over{X^2}}\ , \eqno(26)$$
where $X$ is defined in eq.~(18).
This approximate formula works amazingly well
throughout the full region $0 < \xi < 2$,
and not only in the vicinity of $\xi=2$ [18],
where $X$ symbolizes the radius of the bubble.

Beside the negative mode,
eq.~(25) allows for a discrete positive mode,
also exhibited in Fig.~(9).
For small values of $\xi$ the positive eigenvalue $\omega^2$
increases almost parallel to the continuum
defined by
$$\omega^2={{4\lambda} \over {g^2}}
  \Bigl( {T\over{T_c}}^2-1 \Bigr)\ . $$
When $\xi$ approaches the limiting value of 2, however,
$\omega^2$ bends away from the continuum line.
The positive mode of the bubble now approaches
the positive mode of the antikink
$$\psi(x)= {{{\rm sinh}\Bigl( {1\over{\sqrt{2 \lambda}}}
 {{\delta T_b}\over{M_W}} (x-X) \Bigr) }
 \over{{\rm cosh}^2\Bigl( {1\over{\sqrt{2 \lambda}}}
 {{\delta T_b}\over{M_W}} (x-X) \Bigr) }}\ , \ \ \ \
  \omega^2=3 \Bigr( {{ \delta T_b v} \over {M_H M_W}}\Bigr)^2 \ . $$

For the 3-channel of the bubble the equations do not decouple.
Our analysis indicates, however,
that there are no discrete modes in this channel.
\vskip 0.5 true cm

\noindent {\bf 5) \souligne {Conclusion}}

In this paper we have studied the sphaleron and the bubble,
two non-perturbative solutions of the electroweak model
in the region of the electroweak phase transition.
The temperature dependence of the model is taken into account
via an effective potential, which assures a first order
phase transition.

The sphaleron solution, its action density and
in particular its three-dimensional action
follow an approximate
scaling behaviour with respect to the
well-known zero temperature sphaleron.
The action decreases with the scaling factor
$\Lambda = \langle \phi(T) \rangle / \langle \phi(0)\rangle$,
while the size increases with this scaling factor.
At the temperature $T_c$ the sphaleron is bigger
by a factor of about 2 and at $T_b$ it is bigger by
a factor of about 3 than the zero temperature sphaleron,
giving it a size of about $10-20 M_W$ in the region of the
phase transition.

The sphaleron has a single unstable mode in the region of the
phase transition.
The negative eigenvalue also shows an approximate scaling
behaviour.
Indeed, the eigenvalue $\omega^2$ of the negative mode
at finite temperature
is rather well approximated by the
negative eigenvalue at zero temperature
scaled by the factor $\Lambda^2$.

The bubble solution only exists for sufficiently large temperatures,
$T_c < T < T_b$.
In the limit $T \rightarrow T_c$
the solution converges to the trivial solution
(i.~e.~all fields are vanishing).
In the limit $T \rightarrow T_b$
the bubble solution approaches the antikink solution,
where the bubble wall moves towards infinity,
and the three-dimensional action diverges.

In the vicinity of the phase transition the
bubble function develops an interior region
where it is approximately constant, followed by the region of
the bubble wall.
At temperatures where the phase transition occurs,
the size of both regions is of the same order of magnitude.
The thin wall approximation is not justified here.
It is justified only for $T \rightarrow T_b$,
where the size of the interior region rapidly increases,
while the wall approaches an asymptotic shape and size,
given by the shape and size of the antikink.
The size of the bubble wall,
being a few times the size of the sphaleron,
is thus of the same order of magnitude
as the size of the sphalerons at these temperatures.

The bubble has a single negative mode, the breathing mode.
The frequency of this mode starts from zero at $T_c$
and again approaches zero at $T_b$.
For $T \rightarrow T_b$ the negative mode
of the bubble and its eigenvalue approach the
zero mode of the antikink.
The negative eigenvalue is very well approximated by the simple formula
$\omega^2=-2/X^2$, where $X$ is defined as the value,
where the bubble function changes most strongly,
i.~e.~$|L'(X)|$ is a maximum.
For large temperatures, $X$ may be viewed as the bubble radius.

The bubble also possesses a positive discrete mode.
As a function of temperature
the eigenvalue of this mode increases parallel to the
continuum.
However, for $T \rightarrow T_b$ the eigenvalue
bends away from the continuum:
the positive mode
of the bubble approaches the
positive mode of the antikink.
\vfill\eject

\noindent {\bf 6) \souligne {References}}
\item {[1]} V.A. Kuzmin, V.A. Rubakov and M.E. Shaposhnikov, Phys. Lett.
 B155 (1985) 36.
\item {[2]} F.R. Klinkhamer and N.S. Manton, Phys. Rev. D30 (1984) 2212.
\item {[3]} P. Arnold and L. McLerran, Phys. Rev. D36 (1987) 581;
 D37 ( 1988) 1020.
\item {   } L. Carson and L. McLerran, Phys. Rev. D41 (1990) 647.
\item {   } L. Carson, X. Li, L. McLerran and R.-T. Wang,
 Phys. Rev. D42 (1990) 2127.
\item {[4]} M.E. Shaposhnikow, Nucl. Phys. B287 (1987) 757;
 B299 (1988) 797;
 ``Anomalous Fermion Number Non-Conservation'' CERN-TH. 6304/91 (1991).
\item {[5]} E.W. Kolb and M.S. Turner, ``The Early Universe'',
 Addison-Wesley Publishing Company, Redwood City, 1990.
\item {[6]} S. Coleman, Phys. Rev. D15 (1977) 2929.
\item {   } C. Callan and S. Coleman, Phys. Rev. D16 (1977) 1762.
\item {[7]} A. Linde, Nucl. Phys. B216 (1983) 421.
\item {[8]} M.E. Carrington, Phys. Rev. D45 (1992) 2933.
\item {[9]} M. Dine, R.G. Leigh, P. Huet, A. Linde and D. Linde,
 Phys. Rev. D46 (1992) 550.
\item {[10]} N. Turok, Phys. Rev. Lett. 68 (1992) 1803.
\item {[11]} B.H. Liu, L. McLerran and N. Turok,
 Phys. Rev. D46 (1992) 2668.
\item {[12]} L. Dolan and R. Jackiw, Phys. Rev. D9 (1974) 3320.
\item {[13]} S. Braibant, Y. Brihaye and J. Kunz,
 ``Sphalerons at Finite Temperature'', THU-93/01 (1993).
\item {[14]} F.C. Adams, ``General solutions for tunneling
 of scalar fields with quartic potentials'', preprint (1993).
\item {[15]} T. Akiba, H. Kikuchi and T. Yanagida,
 Phys. Rev. D40 (1989) 588.
\item {[16]} L. Yaffe, Phys. Rev. D40 (1989) 3463.
\item {[17]} Y. Brihaye and J. Kunz, Phys. Lett. B249 (1990) 90.
\item {[18]} D.E. Brahm, ``Complex effective potentials and critical
 bubbles'', CALT-68-1797 (1992).
\vfill\eject

\noindent {\bf 7) \souligne {Figure captions}}

\item {Fig. 1} The gauge field function $G(x)$
and the Higgs field function $L(x)$ of the sphaleron are shown
as a function of the dimensionless coordinate $x=M_Wr$
for $T=0$ (dashed curves) and $T=T_c$ (solid curves).

\item {Fig. 2} The action density $\sigma(x)$ of the sphaleron
is shown as a function of the coordinate $\Lambda x$
for the temperatures $T=0$ (dashed), $T=T_c$ (dotted)
and $T=T_b$ (solid).
$\Lambda$ is the scaling factor
$\langle \phi(T) \rangle / \langle \phi(T_b)\rangle$.

\item {Fig. 3} The action $S(T)$ of the sphaleron and of the
bubble in units of $M_W/\alpha_w$
are shown as functions of temperature for $T_c \le T \le T_b$,
expressed in terms of the variable $\xi$ for $0 \le \xi \le 2$
(solid).
For comparison the approximate scaling values for the
sphaleron action are also shown (dashed).

\item {Fig. 4} The Higgs field function $L(x)$ is shown for the bubble
as a function of the dimensionless coordinate $x=M_Wr$
for the temperatures $\xi=1.3$ (dotted), $\xi=1.8$ (solid)
and $\xi=1.9$ (dashed).

\item {Fig. 5} The action density $\sigma(x)$ is shown for the bubble
as a function of the dimensionless coordinate $x=M_Wr$
for the temperatures $\xi=1.3$ (dotted), $\xi=1.8$ (solid)
and $\xi=1.9$ (dashed).

\item {Fig. 6} The coordinate $X$, where the bubble derivative
$L'(x)$ has its maximal absolute value,
and the quantity $D=-L(0)/L'(X)$
are shown as functions
of the temperature expressed in terms of $\xi$.

\item {Fig. 7} The quantity $L(0)$ is shown for the bubble as a function
of the temperature expressed in terms of $\xi$
and compared with the minimum of the potential
$\langle \phi(T) \rangle / v$.

\item {Fig. 8} The eigenvalue of the negative mode of the sphaleron
is shown as a function
of the temperature expressed in terms of $\xi$ (solid).
For comparison the approximate scaling values for the
eigenvalue are also shown (dashed).

\item {Fig. 9} The eigenvalues of the discrete modes of the bubble
are shown as functions
of the temperature expressed in terms of $\xi$ (solid).
For comparison
the approximate formula $\omega^2=-2/X^2$ (dashed),
and the continuum (dotted) are also shown.
\vfill\end
$\omega^2=-2/X^2$ (dashed),
and the continuum (dotted) are also shown.
\vfill
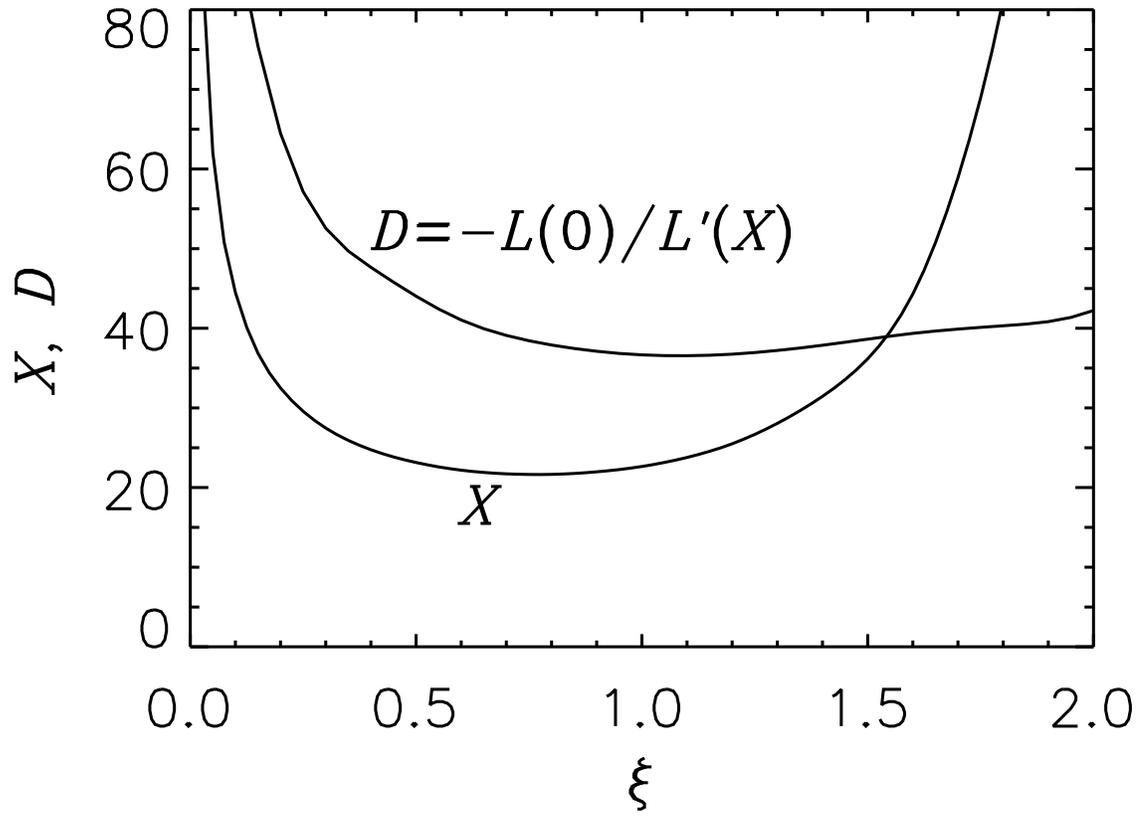
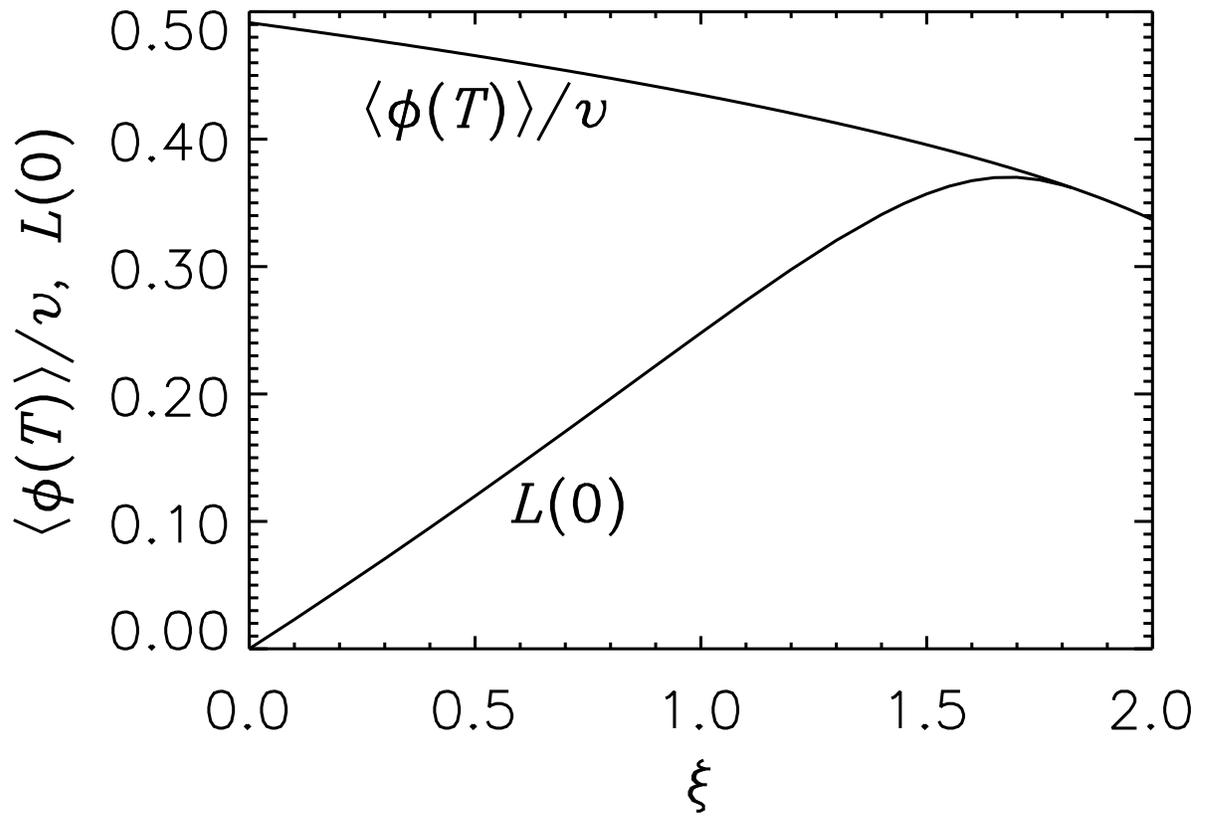
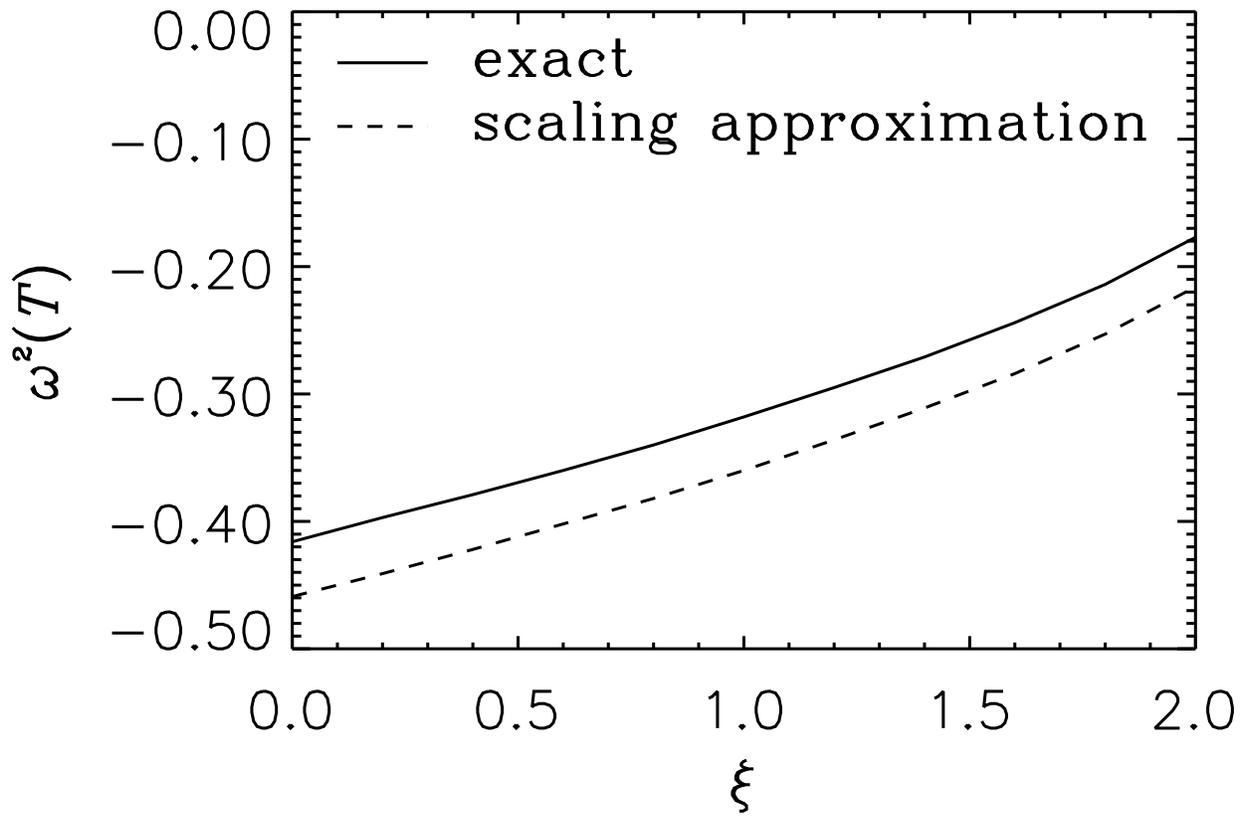
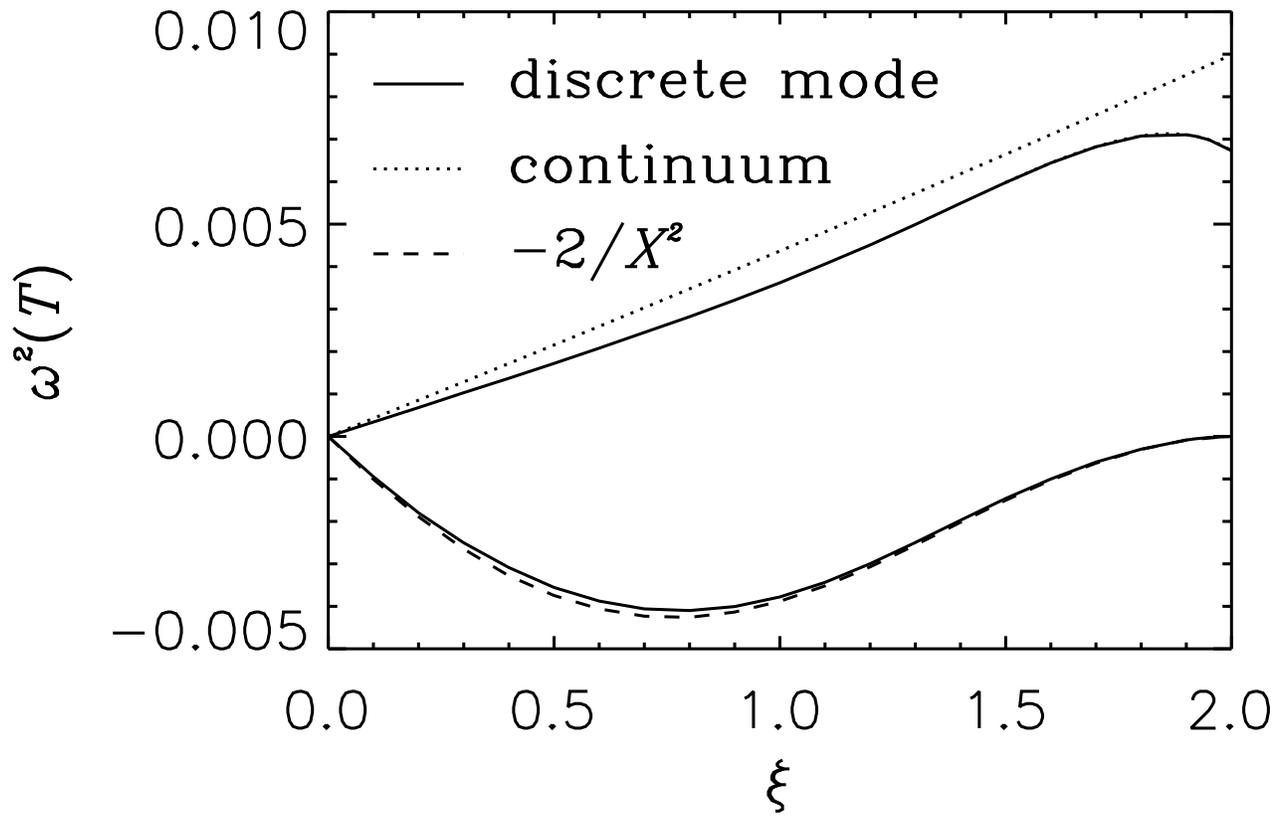
\end